**Multiferroic properties of uniaxially compressed orthorhombic HoMnO$_3$ thin films**

K. Shimamoto,[1,] Y. W. Windsor,[2] Y. Hu,[1] M. Ramakrishnan,[2] A. Alberca,[2] E. M. Bothschafter,[2] L. Rettig,[2] Th. Lippert,[1, 3] U. Staub,[2] and C. W. Schneider[1]

[1)] *Energy and Environment Research Department, Paul Scherrer Institut, CH 5232 Villigen-PSI, Switzerland*

[2)] *Swiss Light Source, Paul Scherrer Institut, CH 5232 Villigen-PSI, Switzerland*

[3)] *Department of Chemistry and Applied Biosciences, Laboratory of Inorganic Chemistry, ETH Zurich, CH 8093 Zurich, Switzerland*

Multiferroic properties of orthorhombic HoMnO$_3$ (*Pbnm* space group) are significantly modified by epitaxial compressive strain along the *a*-axis. We are able to focus on the effect of strain solely along the *a*-axis by using an YAlO$_3$ (010) substrate, which has only a small lattice mismatch with HoMnO$_3$ along the other in-plane direction (the *c*-axis). Multiferroic properties of strained and relaxed HoMnO$_3$ thin films are compared with those reported for bulk, and are found to differ widely. A relaxed film exhibits bulk-like properties such as a ferroelectric transition temperature of 25 K and an incommensurate antiferromagnetic order below 39 K, with an ordering wave vector of (0 $q_b$ 0) with $q_b \approx 0.41$ at 10 K. A strained film becomes ferroelectric already at 37.5 K and has an incommensurate magnetic order with $q_b \approx 0.49$ at 10 K.

Heteroepitaxial growth of oxide materials has been an attractive topic for the past decades opening new research fields in electronics and spintronics. Upon deposition, thin films are forced to grow on their substrate in an energetically favorable manner for the system as a whole, so epitaxial strain is introduced. Since the interatomic distances in the material are then altered, the physical properties of a film can differ from those of bulk. When a film exceeds a certain threshold thickness, called the critical thickness, it exhibits lattice relaxation by introducing defects that allow the lattice constants to approach bulk values.[1–5] Many physical properties can be altered by epitaxial strain, including the highly desirable "ferroic" orderings (memory effects), where the interatomic distances can play a crucial role. For example, the transition temperature of ferromagnetic and ferroelectric (FE) ordering can drastically change, as can the magnitude of the associated order parameters.[6–9] Furthermore, when epitaxial strain is applied to a material which is close to a magnetic and/or electric phase boundary, the ordering motif itself can be changed.[10–12]

The perovskite orthorhombic rare-earth manganites ($o$-$RE$MnO3, $RE$ = Tb - Lu, Y, $Pbnm$ space group) are multiferroics with strong magnetoelectric coupling since they exhibit spin-induced FE.[13,14] This series of materials can be roughly categorized according to two well-known ground states: a $bc$-cycloidal magnetic phase with a $c$-axis FE polarization, and an E-type magnetic phase whose FE polarization is expected to be along the $a$-axis.[15] $o$-HoMnO3 ($o$-HMO) and $o$-YMnO3 ($o$-YMO) sit at the boundary between these phases, exhibiting different magnetic and electric properties depending on the synthesis method (Ho: Ref. 16–22, Y: Ref. 18, 19, 22–25). This indicates that these two materials are sensitive to perturbations. There are several reports on the effects of epitaxial strain on $o$-YMO. For example, when grown coherently on a (010) oriented YAlO3 (YAO (010)) substrate, $o$-YMO films exhibit two electric transitions at 40 and 30 K, while bulk exhibits only one at around 30 K.[24] Magnetic and electric properties change significantly when $o$-YMO is grown on SrTiO3 substrates, exhibiting ferromagnetism and a different magnetoelectric coupling than films grown on YAO substrates.[27–29]

In this report, we focus on epitaxial thin films of $o$-HMO ($a$ = 5.2572 Å, $b$ = 5.8354 Å, $c$ = 7.3606 Å)[30] on YAO ($a$ = 5.1796 Å, $b$ = 5.3286 Å, $c$ = 7.3706 Å)[31] (010) substrates, which have an in-plane lattice mismatch of 1.48 % (compressive) along the *a*-axis and -0.14 % (tensile) along the *c*-axis. Due to the large lattice mismatch between *o*-HMO and YAO, we expect that films will become relaxed when their thickness is sufficiently large. Out study is designed to identify differences in electric and magnetic properties between coherently grown and fully relaxed *o*-HMO.

*o*-HMO films were grown on YAO (010) substrates (Crystec Co. Ltd.) by pulsed laser deposition. Pulsed beams from a KrF excimer laser ($\lambda$ = 248 nm) were focused onto a hexagonal HMO target with a fluence of 2.7 J cm$^{-2}$ at a repetition rate of 2 Hz in a $N_2O$ partial pressure of 0.30 mbar. The substrate temperature was maintained at 780°C by a Si resistive heater with a target-substrate distance of 37 mm. The film thickness was varied from 20 to 400 nm to investigate the effect of epitaxial strain. Lattice parameters of each film were investigated using a four-circle X-ray diffractometer. Capacitance measurements were performed by an Agilent E4980A LCR meter with an AC voltage of 100 mV and FE hysteresis was probed through the Positive-Up Negative-Down (double-wave) method.[32] Magnetic order was probed by resonant soft x-ray diffraction experiments (RSXD) using the RESOXS UHV diffraction end station[33] at the SIM beam line[34] of the Swiss Light Source (SLS). Applied photon energies correspond to the Mn $L_3$ absorption edge ($2p \rightarrow 3d$, ~ 643 eV). Other experimental details are described elsewhere.[35]

The out-of-plane $\theta$ - $2\theta$ scan of a 400 nm film shown in Fig. 1 (a) indicates that the HMO film is in the orthorhombic phase. Reciprocal space maps (RSM) of the (130) and (041) reflection from *o*-HMO thin films with various thicknesses are shown in Fig. 1 (b) and (c), respectively. RSMs from a 20 nm *o*-HMO film clearly demonstrate that the film is coherently grown, and is strained by 1.48 % (compressive) along the *a*- and -0.14 % (tensile) along the *c*-axis. For films thicker than 100 nm, two distinct (130) reflections were observed, demonstrating that the thick *o*-HMO films experience strain relaxation. The critical thickness of an *o*-HMO film on a YAO (010) substrate is around 30 nm for the aforementioned growth condition, considering that the (130) reflection from a 32 nm *o*-HMO thin film starts to broaden towards the bulk position (see Fig. 1 (b)). To roughly illustrate this behavior in the real-space, a schematic cross section of the *o*-HMO thin film along the *a*-axis is shown in Fig. 1 (d). When an *o*-HMO film was grown thicker than the critical thickness, a relaxed layer starts to form on top of the strained one. The width of the diffraction peak of the relaxed layer is larger than the strained layer. This feature indicates that the crystal of a relaxed layer includes larger amounts of defects due to strain relaxation. The RSMs in Fig. 1 (b) and (c) show that the diffraction peaks from the relaxed layer shift towards bulk values as with growing thickness. Furthermore, Fig. 1 (e) presents lattice constants for both layers as functions of thickness. We find that the relaxed layer converges to bulk up to 200 nm, above which the strain is fully relaxed. Here, we note that the shift of the $a$-axis lattice parameter due to strain relaxation is only observed in the relaxed layer. The change in the $c$-axis lattice parameter is small compared to the $a$-axis. It is therefore reasonable to consider that differences between the strained and the relaxed $o$-HMO layers originate from the compressive strain along the $a$-axis.

Electric properties of $o$-HMO films were characterized by using Au (56 nm)/Ti (4 nm) interdigitated electrodes patterned on the film surface (Fig. 2 (a)). The electrodes are aligned with respect to the crystallographic in-plane axis of the film, i.e. $a$- and $c$-axis. Dielectric properties of a 20 nm and a 200 nm $o$-HMO film taken at 15 kHz are shown in Fig. 2. A small dielectric loss (tan $\delta$ < 0.005, Fig. 2 (c), (e)) demonstrates that both 20 and 200 nm samples are well insulating. A larger dielectric loss in the 200 nm film compared to the 20 nm film is due to larger mosaicity in the relaxed film (Fig. 1 (b), (c)) which induces small defect conductance.

The FE transition temperature ($T_{FE}$) of the 20 nm film is 37.5 K as determined from the temperature dependence of the normalized capacitance ($\Delta C(T)$) along the $a$-axis (Fig. 2 (b)). No pronounced transition was observed along the $c$-axis. Unlike in the case of the $o$-YMO thin film on YAO (010),[26] only one transition was observed. A small thermal hysteresis of the $\Delta C(T)$ indicates a weak first order nature of the FE transition (Fig. 2 (b)). These dielectric properties of the strained $o$-HMO film are different from any of the reported bulk results[16–18] but are close to those of $o$-REMnO3 with smaller RE ions such as Tm[36] and Lu.[37]

The 200 nm film exhibited different dielectric properties compared to the strained 20 nm film, as shown in Fig. 2 (d), (e). The $\Delta C(T)$ along the $a$-axis starts to increase at around 42 K, showing a small hump at ≈ 37 K which corresponds to the FE transition of the strained layer (*c.f.* Fig. 1 (d)). It then exhibits a rounded peak and thermal hysteresis, with a crossing between cooling and heating measurements at around 21 K. This behavior is similar to that of polycrystalline bulk samples reported by Lorenz *et al.*[18] The $\Delta C$ along the *c*-axis exhibits a smaller response to the change of temperature than the *a*-axis. The FE transition of the relaxed layer is not straightforward to interpret since $\Delta C(T)$ does not show a clear divergent behavior which is an indication of a FE transition in the conventional Landau theoretical framework.

Ferroelectric hysteresis curves were measured upon heating, settling at each measurement temperature. The effective polarization ($P_{\text{eff}}$) was calculated as $P_{\text{eff}} = Q(tL)^{-1}$, where $Q$ is the measured charge, $t$ is the film thickness, and $L$ is the total length of the finger pairs.[38–40] The effective remnant polarization ($P_{\text{r-eff}}$) was derived from the measured FE hysteresis curves (e.g. Fig. 3 (a) inset) at each temperature. The temperature dependent $P_{\text{r-eff}}$ of the 20 nm film probed by a poling field ($E_{\text{pol}}$) of 47 kV cm$^{-1}$ appears below $T_{\text{FE}}$, and is as large as ~130 nC cm$^{-2}$ at low temperatures. The drop in remnant polarization below ~20 K is caused by the limited input voltage experimentally available which is not large enough to fully polarize the sample.

The FE properties of the 200 nm film are shown in Fig. 3 (b) and (c). A FE hysteresis curve probed by $E_{\text{pol}}$ = 47 kV cm$^{-1}$ from a 200 nm film measured along the *a*-axis at 10 K exhibits a superposition of two FE components (Fig 3 (b) inset). One component is attributed to the relaxed and the other to the strained layer (*c.f.* Fig. 1 (d)). In order to understand the FE transition of the relaxed layer, we focus on the temperature dependence of $P_{\text{r-eff}}$ taken with $E_{\text{pol}}$ = 15 kV cm$^{-1}$ (Fig. 3 (c)) which is immediately above the onset of the $P_{\text{r-eff}}$ v.s. $E_{\text{pol}}$ curve at 10 K (Fig. 3 (c) inset). $P_{\text{r-eff}}(T)$ initially appears below 37 K, which corresponds to the $T_{\text{FE}}$ of the strained layer (Fig. 2 (b)), but disappears below 32 K, indicating that the electric field is not large enough to pole the strained layer. A second onset of $P_{\text{r-eff}}(T)$ appears at 25 K ($T_{\text{FE}}$ of the relaxed layer) which corresponds to the temperature at which the slope of $\Delta C(T)$ becomes steepest (Fig. 3 (d)). This feature was also observed in bulk polycrystalline *o*-HMO by Lorenz et al.[18,19] and its physical origin is still under debate. The $P_{\text{r-eff}}(T)$ shows a jump at 21 K where $\Delta C$ shows a small hump, indicating a two-step FE transition in the relaxed o-HMO layer. Those transitions also have a first order nature exhibited through a thermal hysteresis in $\Delta C(T)$ (Fig. 2 (d)). The orientation of $P_{\text{eff}}$ of the relaxed layer (along the a-axis) is different from that of bulk single crystals reported along the c-axis.[16] This difference may be due to an interfacial interaction between the relaxed and the strained layer, or that a-axis polarization is an intrinsic property, as was theoretically predicted.[41]

The large alteration in $T_{\text{FE}}$ of the strained layer implies change in magnetic properties. From the magnetic resonant diffraction we figured out that the ordering of Mn spins also differs significantly between the strained and the relaxed o-HMO films. Fig. 4 presents the (0 $q_b$ 0) magnetic reflection measured at the Mn $L_3$ from a 32 nm (strained) and a 120 nm (relaxed) o-HMO film. The correlation length[42] calculated from FWHM of the (0 $q_b$ 0) reflection is ~32 nm which is in the order of the probe depth. Since the critical thickness is around 30 nm, we can argue that the measurement on the 120 nm film probes the magnetic properties of the relaxed (top) layer only. The integrated intensities of the (0 $q_b$ 0) reflection as functions of temperature (results not shown) indicates that the Néel temperature of the Mn magnetic order of the strained and the relaxed layers are 41 K and 39 K, respectively. The influence of compressive strain along the a-axis is reflected in the magnitude of $q_b$, presented in Fig. 4 as function of temperature. The strained o-HMO film exhibits a much larger $q_b$ for the whole measured temperature range compared to that of the relaxed o-HMO which corresponds to bulk values.[16,21] Thus, the compressive strain modifies the magnetic modulation vector (0 $q_b$ 0), pushing $q_b$ closer to commensurate E-type ordering (from $q_b \approx 0.41$ to $\approx 0.49$ at ~ 10 K). However, pure commensurate magnetic ordering was not observed from the (0 $q_b$ 0) reflection in agreement with reports for other o-REMnO3 thin films.[35,43,44]

It is thus demonstrated that the compressively strained o-HMO films acquire two key differences in their multiferroic properties with respect to relaxed films and to bulk: an increased value of the $T_{\text{FE}}$, and an enlarged magnetic modulation $q_b$, which is pushed close to the commensurate E-type value of 0.5. It is reasonable to argue that these significant differences are a consequence of the modification of interatomic distances by epitaxial strain. In particular, the $e_g$ state of Jahn-Teller distorted MnO$_6$ octahedra in the *ab*-plane may be strongly influenced since the aspect ratio of the unit cell in the *ab*-plane ($b/a$) is changed by 2.2 %. Considering that the spin ordering of *o-RE*MnO$_3$ is realized by a subtle interplay of exchange interactions, even a small shift of interatomic distances can lead to changes in transition temperatures and stabilize different magnetic ordering schemes.

In summary, we investigated lattice, electric, and magnetic properties of *o*-HMO films grown on YAO (010) substrates. *o*-HMO films grow coherently on the substrate up to a thickness of ~30 nm, and an additional relaxed layer forms on top for thicker films, which exhibits strain relaxation with respect to thickness. The lattice parameters of the relaxed layer converge to near-bulk values when the film is thicker than 200 nm. Electrical characterization and RSXD measurements reveal significant differences in the multiferroic properties of strained and relaxed films. The strained film exhibits $T_{FE}$ = 37.5 K, which is notably higher than bulk and relaxed *o*-HMO film ($\approx$ 25 K). The magnetic ordering vector (0 $q_b$ 0) shifts significantly from $q_b \approx$ 0.41 for bulk and the relaxed films to $\approx$ 0.49 for the strained films. This modification of electric and magnetic ordering induced by the compressive strain along the *a*-axis may be attributed to modulated interatomic distances.

Financial support and CROSS funding to K.S. from PSI are gratefully acknowledged. We would like to thank D. Marty (PSI, Laboratory for Micro- and Nanotechnology) for support with optical lithography and U. Greuter for the implementation of a Sawyer-Tower circuit for ferroelectric hysteresis measurements. The use of the LCR-meter belonging to the Laboratory for Scientific Developments and Novel Materials is acknowledged. Y.W.W. and M.R. are supported through Grants from the Swiss National Science Foundation No. 200020-159220 and a Sinergia Grant No. 200021-137657, respectively. E.M.B. and L.R. are supported by the NCCR MUST from the Swiss National Science Foundation. E.M.B. acknowledges funding from the European Community's Seventh Framework Program (FP7/2007-2013) under grant agreement n.° 290605 (COFUND: PSI-FELLOW). Soft x-ray experiments were performed at the X11MA beamline at SLS, whose beamline staff we gratefully acknowledge.

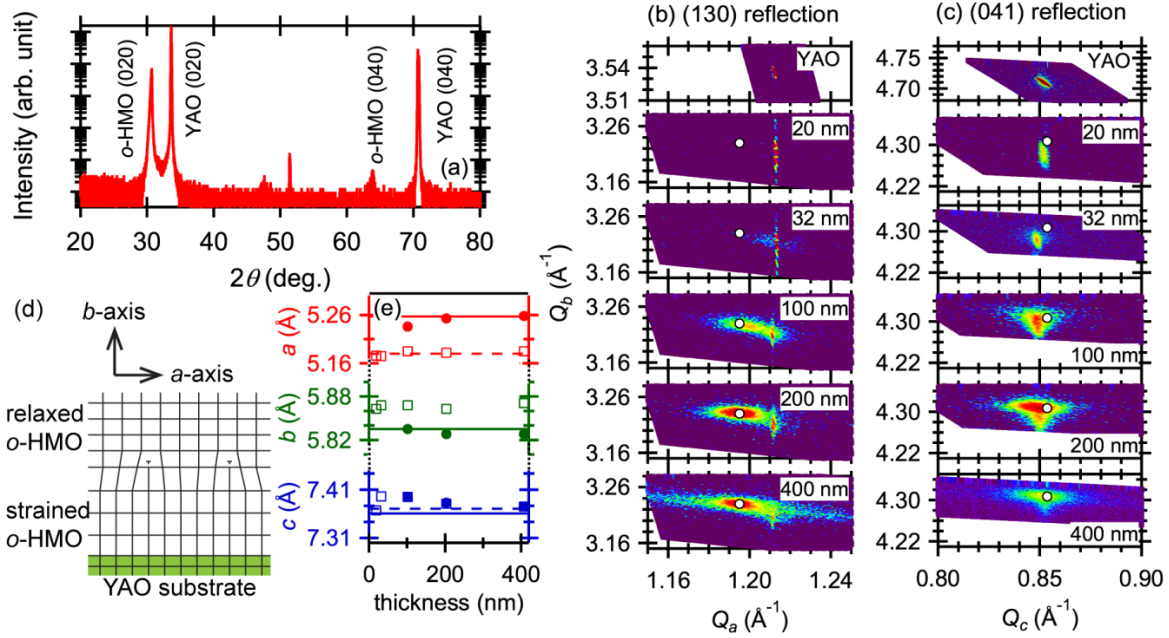

FIG. 1. Lattice parameters of $o$-HMO films grown on YAO (010) substrates. (a) $\theta - 2\theta$ scan of a 400 nm $o$-HMO film. (b), (c) Reciprocal space maps around the (130) and (041) reflections (respectively) of $o$-HMO films of various thicknesses. A white marker in each map indicates the reflections' position for bulk $o$-HMO. (d) Schematic side-view of the relaxed and the strained layers of the film. (e) Lattice parameters of $o$-HMO films derived from (b) and (c). Open symbols are those of the strained layer and closed ones are of the relaxed layer. Horizontal solid and dashed lines indicate the values of $o$-HMO[30] and of YAO[31] respectively.

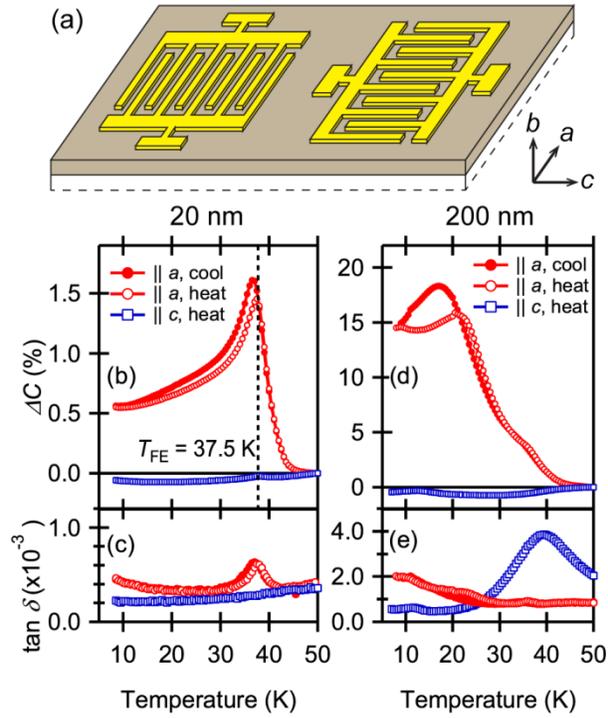

FIG. 2. (a) A schematic image of interdigitated electrodes on a thin film. Temperature dependent electric properties of (b) (c) the 20 nm and (d) (e) the 200 nm $o$-HMO films respectively: (b) (d) normalized capacitance and (c) (e) loss tangent. The normalized capacitance is derived by $\Delta C = (C(T) - C(50\ \text{K}))/C(50\ \text{K})$.

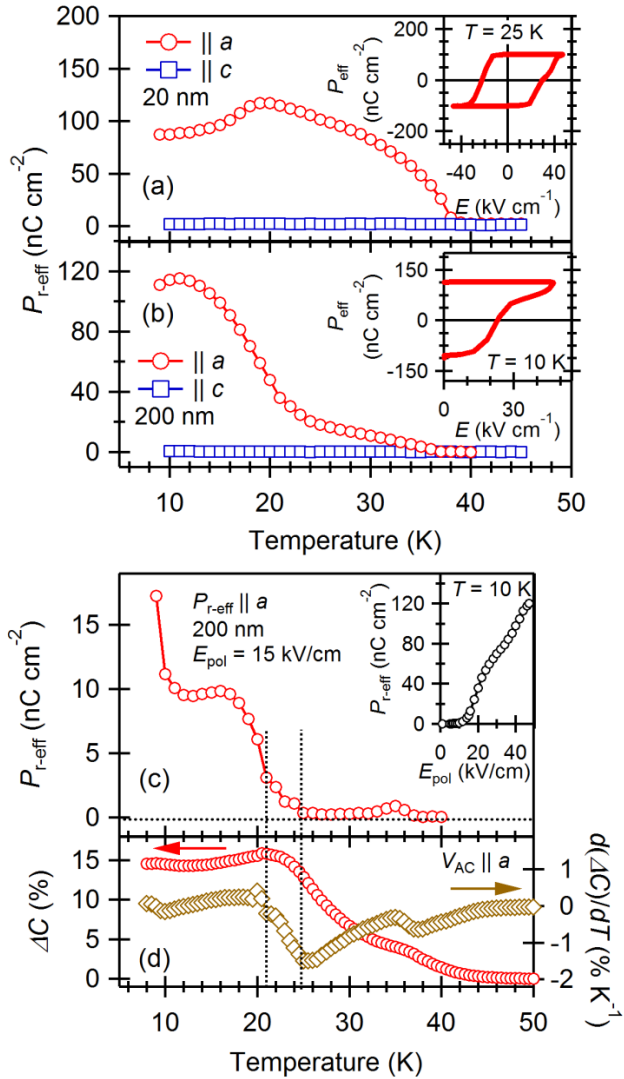

FIG. 3. Temperature dependent $P_{\text{r-eff}}$ of the 20 nm (a) and the 200 nm (b) $o$-HMO film derived from FE hysteresis curves measured using $E_{\text{pol}} = 47$ kV cm$^{-1}$. A part of the FE hysteresis curve is shown in the inset of each figure. (c) The $P_{\text{r-eff}}(T)$ of the 200 nm $o$-HMO film derived from FE hysteresis curves measured using $E_{\text{pol}} = 15$ kV cm$^{-1}$. The $P_{\text{r-eff}}$ at 10 K is plotted as a function of $E_{\text{pol}}$ in the inset. (d) The normalized capacitance and its derivative with respect to temperature along the $a$-axis.

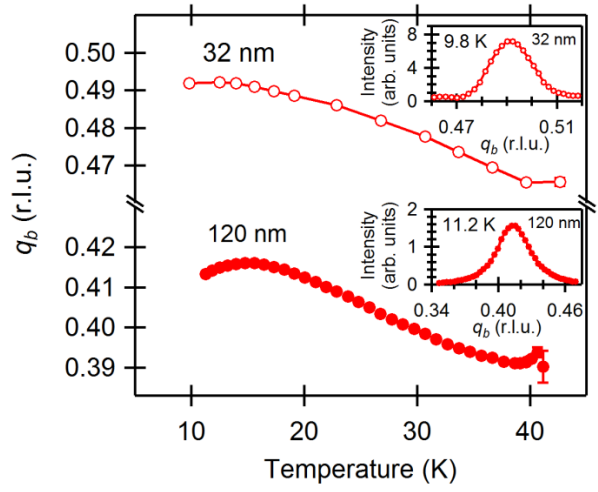

FIG. 4. Temperature dependence of the magnetic modulation parameter $q_b$ derived from $(0\ q_b\ 0)$ reflections of a 32 nm (open symbols) and a 120 nm (closed symbols) thin film. The $(0\ q_b\ 0)$ reflection at a selected temperature is shown in the insets.